\documentclass[aps,pre,twocolumn,groupedaddress]{revtex4-1}

\usepackage{amsmath,amssymb}
\usepackage{color}
\usepackage{graphicx}
\usepackage{bm}

\begin{document}


\title{Duality-based calculations for transition probabilities in stochastic chemical reactions}


\author{Jun Ohkubo}
\affiliation{
Graduate School of Science and Engineering, Saitama University,
255 Shimo-Okubo, Sakura, Saitama, 338-8570, Japan
}



\begin{abstract}
An idea for evaluating transition probabilities
in chemical reaction systems is proposed,
which is efficient for repeated calculations with various rate constants.
The idea is based on duality relations;
instead of direct time-evolutions of the original reaction system,
the dual process is dealt with.
Usually, if one changes rate constants of the original reaction system,
the direct time-evolutions should be performed again,
using the new rate constants.
On the other hands,
only one solution of an extended dual process 
can be re-used to calculate the transition probabilities
for various rate constant cases.
The idea is demonstrated in a parameter estimation problem
for the Lotka-Volterra system.
\end{abstract}



\maketitle


\section{Introduction}
\label{sec_intro}

Recent developments of experimental techniques
enable us to obtain detailed data for bio-chemical reactions
within cells,
and the importance of the role of data analysis has been increased.
In small systems such as cells,
it has been shown that discrete characteristics could play important roles 
\cite{Rao2002},
and hence chemical reactions in such small systems
should be treated as chemical master equations.
That is, conventional rate equations are not adequate, 
in which the noise effects are neglected and 
numbers of chemical substances are approximated 
as continuous variables.
Hence, it would be necessary to treat discrete variables directly.
The time-evolution of the discrete variables is directly
expressed via the chemical master equations.
In the past, various analytical and numerical methods for the chemical master equations
have been developed (for example, see \cite{Gardiner_book}.)

Transition probabilities in the chemical master equations are 
one of the important quantities,
especially in time-series data analysis.
For example, consider an estimation problem for rate constants
from time-series data of chemical reactions.
When all time-series data for numbers of chemical substances and chemical reactions are available,
the Bayesian statistics gives easily the estimation of the rate constants \cite{Wilkinson_book}.
However, it is in general difficult to obtain such detailed time-series data,
and partial observations with discrete time steps should be considered in realistic experiments.
In general, 
the partial observation cases need many numerical time-evolutions for different rate constants,
and large computational costs are needed.
For example, in \cite{Wang2010}, an estimation method
based on gradient descent techniques has been proposed,
and the method needs many iterated calculations to seek
rate constants with the largest likelihood value.
Although sometimes the estimation problems have been performed
via approximations \cite{Golightly2005,Ruttor2009,Vrettas2010},
these approximations treat discrete variables as continuous ones,
and then the approximations could be inadequate for small systems such as cells.
Hence, it is useful to develop 
more rapid and direct evaluation methods for transition probabilities
for different rate constant cases.

In the present paper,
an idea to evaluate transition probabilities in chemical master equations is proposed.
The idea is based on duality concepts;
instead of direct time-evolutions for the original chemical master equations,
the corresponding dual processes are evaluated.
Employing extensions of states,
it is possible to construct an extended dual process
which does not explicitly include some rate constants of the original processes.
Furthermore, it will be shown that only a time-evolution for the extended dual process
enables us to evaluate transition probabilities for the original chemical master equations
with \textit{various} rate constants.
This characteristics of the extended dual processes
could reduce largely the computational costs for 
the calculations of the transition probabilities
for various rate constants. 

This paper is organized as follows.
In Sec.~\ref{sec_general}, 
the basic idea for the usage of the dual process is explained.
Section~\ref{sec_demo} gives a demonstration
for the derivation of the dual processes
using the Lotka-Volterra system.
Concluding remarks are given in Sec.~\ref{sec_conclusion}.

\section{Duality relations for transition probabilities}
\label{sec_general}

In this section, the idea based on the duality relations is explained. 
A concrete example for the derivations of the dual process
will be given in Sec.~\ref{sec_demo}.

\subsection{Stochastic chemical reaction systems}

Consider a stochastic chemical reaction system
with $M$ chemical substances, $X_1, \dots, X_M$.
Denote $\bm{n} = (n_1, \dots, n_M)$ as the state of the reaction systems,
where $n_m$ denotes the number of chemical substance $X_m$.
Note that $n_m$ takes any positive integer or zero ($n_m \in \{0, 1, 2, \cdots\}$.)
Assume that there are $R$ chemical reactions,
which take the form 
\begin{align}
\begin{cases}
u_{11} X_1 + \cdots u_{1M} X_M \xrightarrow{c_1} v_{11} X_1 + \cdots v_{1M} X_M, \\
u_{21} X_1 + \cdots u_{2M} X_M \xrightarrow{c_2} v_{21} X_1 + \cdots v_{2M} X_M, \\
\qquad  \vdots \\
u_{R1} X_1 + \cdots u_{RM} X_M \xrightarrow{c_R} v_{R1} X_1 + \cdots v_{RM} X_M,
\end{cases}
\end{align}
where $u_{rm}$ and $v_{rm}$ correspond
to the stoichiometries associated with the $m$-th reactant and product
of the $r$-th reaction, respectively.
The $r$-th chemical reaction has a rate constant $c_r$.
The actual rate for the $r$-th reaction depends on the number of chemical substances,
and a rate law or hazard, $h_r(\bm{n}, c_r)$, is introduced \cite{Wilkinson_book}.
For example, for $\displaystyle 2 X_j + X_k \xrightarrow{c} X_{l}$,
the hazard should be defined as $\displaystyle h(\bm{n},c) = c \frac{n_j (n_j-1)}{2} n_k$,
and so on.
In the present paper, the following redefined hazard, $h'_r(\bm{n})$, is introduced:
\begin{align}
h_r(\bm{n},c_r) &= \tilde{c}_r h'_r(\bm{n}),
\end{align}
where
\begin{align}
\tilde{c}_r \equiv c_r \prod_{m=1}^M \frac{1}{u_{rm}!},
\end{align}
and
\begin{align}
h'_r(\bm{n}) \equiv  \prod_{m=1}^M \frac{n_m!}{(n_m-u_{rm})!}. 
\end{align}
These notations are convenient to describe the extension of the dual process later.
In addition, using the quantities $\{u_{rm}\}$ and $\{v_{rm}\}$,
the net effect reaction matrix $A$ is defined \cite{Golightly2005};
the components of $A$, $\{a_{rm}\}$, are defined as $a_{rm} = v_{rm} - u_{rm}$.

Using the above notations,
the chemical master equations are written as follows:
\begin{align}
&\frac{\partial}{\partial t} P(\bm{n},t) \nonumber \\
&= \sum_{r=1}^R
\left[
\tilde{c}_r h'_r(\bm{n}-A_r) P(\bm{n}-A_r,t) - \tilde{c}_r h'_r(\bm{n}) P(\bm{n},t)
\right],
\label{eq_master}
\end{align}
where $A_r$ is the $r$-th row of the matrix $A$.
For details of the master equations,
see, for example, \cite{Wilkinson_book,Gardiner_book}.

When one employs the direct numerical time-integration 
for the original chemical master equations in Eq.~\eqref{eq_master}
(with a suitable truncation for finite numbers of equations),
the transition probability for a certain initial state $\bm{n}$
to final state $\bm{m}$ with fixed rate constants $\{c_1, \dots, c_R\}$
can be evaluated.
If one wants to know the transition probabilities
for different rate constant cases,
the numerical time-integrations must be performed again with the new rate constants;
these repeated time-integrations are time-consuming
when one wants to know transition probabilities for various rate constant cases.
Hence, in the following discussions,
the chemical master equations are investigated 
from the viewpoint of bosonic operators,
which enables us to obtain an idea to avoid
the repeated time-integrations.

\subsection{Doi-Peliti formalism}

The chemical master equations in Eq.~\eqref{eq_master}
are essentially the infinite number of coupled ordinary
differential equations.
There are some analytical methods to rewrite the chemical master equations,
and one of the methods is the so-called Doi-Peliti formalism
\cite{Doi1976a,Doi1976b,Peliti1985}.
The Doi-Peliti formalism has been widely used,
ranging from the research fields of reaction-diffusion systems \cite{Tauber2005}
to genetic switches \cite{Sasai2003,Mugeler2009}.
The method is based on the algebraic probability theory \cite{Hora_book,Ohkubo2013a},
and the following bosonic creation operators for the $m$-th chemical substance, 
$a^\dagger_m$, and annihilation operators, $a_m$, are used:
\begin{align}
&[ a_m, a^\dagger_m] \equiv a_m a^\dagger_m - a^\dagger_m a_m = 1, \\ 
&[a^\dagger_m,a^\dagger_m] = [a_m,a_m] = 0, \\
&[a_{m}, a^\dagger_{m'}] = [a^\dagger_{m}, a_{m'}] \nonumber \\
&\quad = [a_{m}, a_{m'}] = [a^\dagger_{m},a^\dagger_{m'}] = 0,
\textrm{ for $m \neq m'$}.
\end{align}
That is, the creation and annihilation operators
for the same chemical substance $X_m$ do not commute with each other;
for different chemical substances, these operators can commute.

The actions of the creation and annihilation operators on state 
$| \bm{n} \rangle = | n_1, \dots, n_M \rangle$
in a Fock space are defined as
\begin{align}
&a^\dagger_m |n_1, \dots, n_m, \dots, n_M \rangle 
= |n_i, \dots, n_m+1, \dots, n_M \rangle, \\
&a_m |n_1, \dots, n_m, \dots, n_M \rangle 
= n_m |n_1, \dots, n_m-1, \dots, n_M \rangle.
\end{align}
The corresponding dual (bra) states $\langle \bm{m} | = \langle m_1, \dots, m_M |$
are introduced
as satisfying the following inner product:
\begin{align}
\langle \bm{m} | \bm{n} \rangle = \delta(n_1,m_1) \cdots \delta(n_M,m_M) \bm{n}!,
\label{eq_inner_product}
\end{align}
where $\delta(n_m,m_m)$ is the Kronecker delta,
and
\begin{align}
\bm{n}! \equiv n_1! \cdots n_M!.
\end{align}
It is easy to confirm that the actions of 
the creation and annihilation operators to the bra states
become as follows:
\begin{align}
&\langle n_1, \dots, n_m, \dots, n_M | a^\dagger_m = n_m \langle n_1, \dots, n_m-1, \dots, n_M |,  \\
&\langle n_1, \dots, n_m , \dots, n_M | a_m = \langle n_1, \dots, n_m+1, \dots, n_M |.
\end{align}

Using the above notations,
the state $| P(t) \rangle$ is defined as
\begin{align}
\left| P(t) \right\rangle = \sum_{n_1=0}^\infty \dots \sum_{n_M=0}^\infty
P(\bm{n},t) | \bm{n} \rangle.
\end{align}
In order to derive the time-evolution equation
for the state $|P(t) \rangle$,
the following quantities are introduced,
which correspond to the redefined hazards in the original chemical master equations:
\begin{align}
&\mathcal{H}_r = \prod_{m=1}^M \left( a^\dagger_m \right)^{v_{rm}} \left( a_m \right)^{u_{rm}}, \\
&\mathcal{H}_r^\mathrm{diag} = \prod_{m=1}^M \left( a^\dagger_m \right)^{u_{rm}} \left(  a_m\right)^{u_{rm}}.
\end{align}
Then, the original chemical master equations in Eq.~\eqref{eq_master}
are rewritten as follows:
\begin{align}
\frac{d}{dt} | P(t) \rangle = \mathcal{L} | P(t) \rangle,
\label{eq_master_Doi_Peliti}
\end{align}
where
\begin{align}
\mathcal{L} = \sum_{r=1}^R \mathcal{L}_r, 
\end{align}
and
\begin{align}
\mathcal{L}_r = \tilde{c}_r \left( \mathcal{H}_r - \mathcal{H}_r^\mathrm{diag} \right).
\end{align}

Finally, the transition probability $P_{\bm{n}\to \bm{m}}(t)$ is
written in terms of the Doi-Peliti formalism as
\begin{align}
P_{\bm{n} \to \bm{m}}(t) = \frac{1}{\bm{m}!} \langle \bm{m} | P(t) \rangle 
= \frac{1}{\bm{m}!} \langle \bm{m} | e^{\mathcal{L} t} | \bm{n} \rangle,
\label{eq_transition_probability_in_Doi_Peliti}
\end{align}
where $e^{\mathcal{L}t} | \bm{n} \rangle$
corresponds to a solution of the time-evolution
starting from state $\bm{n}$.
Note that we need the factor $1/\bm{m}!$
because of the characteristics of 
the inner product in Eq.~\eqref{eq_inner_product}.

\subsection{Extended dual process}

Here, additional bosonic operators are introduced
to obtain an extended dual process,
in which some rate constants are vanished
and additional states emerge.
For simplicity, here only one rate constant $c_1$
will be replaced with the additional bosonic operator;
extensions for multiple cases are straightforward.

First, additional bosonic operators,
$a^\dagger_{c_1}$ and $a_{c_1}$, are introduced.
Here, the following property of the coherent states is important:
\begin{align}
a_{c_1} | z_1 \rangle = z_1 | z_1 \rangle,
\label{eq_coherent}
\end{align}
where $|z_1 \rangle$ is the coherent state with parameter $z_1 \in \mathbb{R}$,
which is defined as
\begin{align}
| z_1 \rangle \equiv e^{z_1 a_{c_1}^\dagger} | 0 \rangle.
\end{align}
In addition, noting $\langle n_{c_1} | = \langle 0 | (a_{c_1})^{n_{c_1}}$, 
we have
\begin{align}
\langle n_{c_1} | z_1 \rangle = z_1^{n_{c_1}}.
\end{align}

Second, the following time-evolution operator $\mathcal{L}^\mathrm{ex}$
is introduced instead of the original $\mathcal{L}$:
\begin{align}
\mathcal{L}^\mathrm{ex} = \mathcal{L}^\mathrm{ex}_1 + \sum_{r=2}^R \mathcal{L}_r,
\end{align}
where
\begin{align}
\mathcal{L}^\mathrm{ex}_1 = a_{c_1} \left(\mathcal{H}_1 - \mathcal{H}_1^\mathrm{diag} \right).
\end{align}
Furthermore, the state $|P(t)\rangle$ is extended as
\begin{align}
| P^\mathrm{ex}(t)\rangle = \sum_{n_1 = 0}^\infty \cdots \sum_{n_M = 0}^\infty
P(\bm{n},t) | \bm{n} \rangle | z_1 \rangle,
\end{align}
and the time-evolution for the extended state $| P^\mathrm{ex}(t)\rangle$
obeys
\begin{align}
\frac{\partial}{\partial t} | P^\mathrm{ex}(t) \rangle
= \mathcal{L}^\mathrm{ex} | P^\mathrm{ex}(t) \rangle.
\label{eq_extended_time_evolution}
\end{align}
Note that $\mathcal{L}^\mathrm{ex}$ in Eq.~\eqref{eq_extended_time_evolution}
gives the same quantity with $\mathcal{L}$ in Eq.~\eqref{eq_master_Doi_Peliti}
when $z_1 = \tilde{c}_1$,
because of the characteristics of the coherent state in Eq.~\eqref{eq_coherent}.

Third, the following bra state is defined:
\begin{align}
\langle \phi(t) | = 
\sum_{n_{c_1}=0}^\infty \sum_{n_1=0}^\infty \cdots \sum_{n_M=0}^\infty
\phi(\bm{n}, n_{c_1}, t) \langle n_{c_1} | \langle \bm{n} |,
\end{align}
where $\langle \bm{n} | = \langle n_1, \dots, n_M |$,
and $\langle n_{c_1} |$ corresponds to the state
related to the additional bosonic creation and annihilation operators
$a_{c_1}^\dagger$ and $a_{c_1}$.

Fourth, instead of the time-evolution for $|P^\mathrm{ex}(t)\rangle$,
the following time-evolution for the bra state is considered:
\begin{align}
\frac{\partial}{\partial t} \langle \phi(t) |
= \langle \phi(t) | \mathcal{L}^\mathrm{ex}, 
\end{align}
that is,
\begin{align}
\frac{\partial}{\partial t} | \phi(t) \rangle
= \left(\mathcal{L}^\mathrm{ex}\right)^\dagger | \phi(t) \rangle,
\label{eq_dual_time_evolution}
\end{align}
where $\left(\mathcal{L}^\mathrm{ex}\right)^\dagger$
is the conjugate of $\mathcal{L}^\mathrm{ex}$.
Equation~\eqref{eq_dual_time_evolution} is written 
in terms of the Doi-Peliti formalism,
and it is possible to derive the corresponding
infinite coupled ordinary differential equations
for $\{\phi(\bm{n},n_{c_1},t)\}$.
That is, introducing the following `hazard':
\begin{align}
\tilde{h}'_r(\bm{n})
&= \prod_{m=1}^M \frac{n_m!}{(n_m - v_{rm})!},
\end{align}
we have
\begin{align}
&\frac{\partial}{\partial t} \phi(\bm{n},n_{c_1},t) \nonumber \\
&=
\tilde{h}'_1(\bm{n}+A_r) \phi(\bm{n}+A_r,n_{c_1}-1,t) - h'_r(\bm{n}) \phi(\bm{n},n_{c_1}-1,t) \nonumber \\
&+\sum_{r=2}^R
\left[
\tilde{c}_r \tilde{h}'_r(\bm{n}+A_r) \phi(\bm{n}+A_r,n_{c_1},t) - \tilde{c}_r h'_r(\bm{n}) \phi(\bm{n},n_{c_1}, t)
\right].
\end{align}

Here, note that the time-evolution equations
for $\{\phi(\bm{n},n_{c_1},t)\}$
are not the chemical master equations in general;
the equations do not satisfy the probability conservation law,
and then $\{\phi(\bm{n},n_{c_1},t)\}$
cannot be interpreted as a probability distribution.
Although it is possible to recover the probabilistic nature
by using similar discussions
used in \cite{Ohkubo2013b},
it is enough to use $\{\phi(\bm{n},n_{c_1},t)\}$ 
to calculate the transition probabilities here.

The main idea in the present paper
is the replacement of the time-evolution for $|P(t)\rangle$
with $\langle \phi(t)|$.
Hence, 
\begin{align}
&P_{\bm{n}\to \bm{m}}(t) \nonumber \\
&= \frac{1}{\bm{m}!} \langle \bm{m} | e^{\mathcal{L}t} | \bm{n} \rangle \nonumber \\
&= \frac{1}{\bm{m}!} \langle n_{c_1} = 0| \langle \bm{m} | e^{\mathcal{L}^\mathrm{ex}t}
|\bm{n}\rangle |z_1 = \tilde{c}_1 \rangle \nonumber \\
&= \frac{1}{\bm{m}!} 
\sum_{n_{c_1} = 0}^\infty \sum_{n'_1=0}^\infty \dots \sum_{n'_M=0}^\infty 
\phi(\bm{n}',n_{c_1},t)
\langle n_{c_1} | \langle \bm{n}' | \bm{n} \rangle | z_1 = \tilde{c}_1 \rangle \nonumber \\
&= \frac{\bm{n}!}{\bm{m}!} \sum_{n_{c_1}=0}^\infty 
\tilde{c}_1^{n_{c_1}}
\phi(\bm{n},n_{c_1},t),
\label{eq_main_result}
\end{align}
where the initial condition for the bra state is
\begin{align}
&\phi(n'_1, \dots, n'_M, n_{c_1}, t=0) \nonumber \\
&= \begin{cases}
1 & \textrm{for $n'_1 = m_1, \dots, n'_M = m_M, n_{c_1} = 0$}, \\
0 & \textrm{otherwise.}
\end{cases}
\end{align}

The above discussions imply the following fact:
The transition probabilities with the rate constant $c_1$
can be evaluated from Eq.~\eqref{eq_main_result}
using the solutions of the extended dual process.
Note that the solutions for the extended dual process,
$\{ \phi(\bm{n}, n_{c_1}, t) \}$,
does not include the rate constant $c_1$ explicitly.
Hence, only one numerical time-integration
for the extended dual process is necessary to evaluate
the transition probabilities $\bm{n} \to \bm{m}$
for various $c_1$ cases.


\section{Demonstration of the derivation and applications of dual processes}
\label{sec_demo}

Here, a demonstration for the derivation of the extended dual process
is shown by using the famous Lotka-Volterra system, which 
has been already used for parameter estimation problem in \cite{Ruttor2009}:
\begin{align}
\begin{cases}
X_1 \xrightarrow{\alpha} 2 X_1, \\
X_1 + X_2 \xrightarrow{\beta} X_2, \\
X_1 + X_2 \xrightarrow{\delta} X_1 + 2 X_2, \\
X_2 \xrightarrow{\gamma} \emptyset.
\end{cases}
\end{align}

The chemical master equation for the Lotka-Volterra system
is written as
\begin{align}
& \frac{d P(n_1,n_2,t)}{dt} \nonumber \\
&=\alpha \left[ (n_1-1)P(n_1-1, n_2, t) - n_1 P(n_1, n_2, t)\right] \nonumber \\
&\quad + \beta \left[ (n_1+1) n_2 P(n_1+1, n_2, t) - n_1 n_2 P(n_1, n_2, t)\right] \nonumber \\
&\quad + \delta \left[ n_1 (n_2-1) P(n_1, n_2-1, t) - n_1 n_2 P(n_1, n_2, t)\right] \nonumber \\
&\quad + \gamma \left[ (n_2+1) P(n_1, n_2+1, t) - n_2 P(n_1, n_2, t)\right],
\label{eq_demo_master_equation}
\end{align}
where $n_1$ and $n_2$ are the numbers of particles $X_1$ and $X_2$, respectively.
The time-evolution operator in the Doi-Peliti formalism is defined as
\begin{align}
\mathcal{L} =& 
\alpha (a_1^\dagger a_1^\dagger a_1 - a_1^\dagger a_1 ) 
+ \beta(a_1 a_2^\dagger a_2 - a_1^\dagger a_1 a_2^\dagger a_2) \nonumber \\
&+ \delta(a_1^\dagger a_1 a_2^\dagger a_2^\dagger a_2 - a_1^\dagger a_1 a_2^\dagger a_2)
+ \gamma(a_2 - a_2^\dagger a_2).
\label{eq_demo_time_evolution_operator}
\end{align}

We assume that
precise values for the rate constants $\alpha$ and $\beta$ are unknown.
Therefore, the two bosonic operators ($a_\alpha$ and $a_\beta$) 
and their adjoint operators ($a_\alpha^\dagger$ and $a_\beta^\dagger$)
are introduced.
Using certain constants $\alpha'$ and $\beta'$,
$\alpha$ and $\beta$ in Eq.~\eqref{eq_demo_time_evolution_operator}
are replaced as $\alpha' a_\alpha$ and $\beta' a_\beta$, respectively,
and then the following extended time-evolution operator is derived:
\begin{align}
\mathcal{L}^\mathrm{ex} 
=&
\alpha' a_\alpha (a_1^\dagger a_1^\dagger a_1 - a_1^\dagger a_1 ) 
+ \beta' a_\beta(a_1 a_2^\dagger a_2 - a_1^\dagger a_1 a_2^\dagger a_2) \nonumber \\
&+ \delta(a_1^\dagger a_1 a_2^\dagger a_2^\dagger a_2 - a_1^\dagger a_1 a_2^\dagger a_2)
+ \gamma(a_2 - a_2^\dagger a_2).
\label{eq_demo_time_evolution_operator_extended}
\end{align}
Note that the replacement $\alpha \to \alpha' a_\alpha$ is used here,
instead of $\alpha \to a_\alpha$ (and the same as $\beta$).
These replacements do not mean approximations;
they correspond to simple variable transformations.
The reasons to introduce the replacements are as follows:
\begin{itemize}
\item
Sometimes we know the rough values (or only the orders) of unknown rate constants;
these additional information can be embedded into the extended dual process
with small modifications of the discussions in Sec.~II,
as we will see here.
\item
As we will see below (Eq.~\eqref{eq_demo_final_dual}), the final expression
corresponds to the Taylor-type (Maclaurin-type) expansion.
Hence, in practical, it is important to use small $\alpha/\alpha'$ and $\beta / \beta'$
to confirm the rapid convergence of the summation in Eq.~\eqref{eq_demo_final_dual}.
Hence, sometimes the replacements reduce the practical computational issues.
\end{itemize}

The adjoint operator for $\mathcal{L}^\mathrm{ex}$ is
\begin{align}
\left( \mathcal{L}^\mathrm{ex} \right)^\dagger
=&
\alpha' a_\alpha^\dagger (a_1^\dagger a_1 a_1 - a_1^\dagger a_1 ) 
+ \beta' a_\beta^\dagger (a_1^\dagger a_2^\dagger a_2 - a_1^\dagger a_1 a_2^\dagger a_2) \nonumber \\
&+ \delta(a_1^\dagger a_1 a_2^\dagger a_2 a_2 - a_1^\dagger a_1 a_2^\dagger a_2)
+ \gamma(a_2^\dagger - a_2^\dagger a_2),
\end{align}
and using the discussions in Sec.~\ref{sec_general},
the extended dual process obeys the following time-evolution equation:
\begin{align}
\frac{\partial}{\partial t} |\phi(t)\rangle = 
\left( \mathcal{L}^\mathrm{ex} \right)^\dagger
|\phi(t)\rangle,
\label{eq_demo_final_pre}
\end{align}
where
\begin{align}
&|\phi(t)\rangle \nonumber \\
&= 
\sum_{n_\alpha = 0}^\infty
\sum_{n_\beta = 0}^\infty
\sum_{n_1=0}^\infty
\sum_{n_2=0}^\infty
\phi(n_1, n_2, n_\alpha, n_\beta,t) 
|n_1\rangle |n_2\rangle | n_\alpha \rangle | n_\beta \rangle.
\end{align}
Then, the following coupled ordinary differential equations are derived:
\begin{align}
&\frac{\partial}{\partial t} \phi(n_1, n_2, n_\alpha, n_\beta,t)  \nonumber \\ 
&= \alpha' [
(n_1+1) n_1 \phi(n_1+1, n_2, n_\alpha-1, n_\beta,t) \nonumber \\
&\qquad \,\, - n_1 \phi(n_1, n_2, n_\alpha-1, n_\beta,t)
] \nonumber \\
&\quad + \beta' [
n_2 \phi(n_1-1, n_2, n_\alpha, n_\beta-1,t) \nonumber \\
&\qquad \quad \,\, - n_1 n_2 \phi(n_1, n_2, n_\alpha, n_\beta-1,t)
] \nonumber \\
&\quad + \delta [
n_1 (n_2+1) n_2 \phi(n_1, n_2+1, n_\alpha, n_\beta,t) \nonumber \\
&\qquad \quad \, - n_1 n_2 \phi(n_1, n_2, n_\alpha, n_\beta,t)
] \nonumber \\
&\quad + \gamma \left[
\phi(n_1, n_2-1, n_\alpha, n_\beta,t)
- n_2 \phi(n_1, n_2, n_\alpha, n_\beta,t)
\right].
\label{eq_demo_final}
\end{align}
Finally, the solutions of Eq.~\eqref{eq_demo_final}
are used to evaluate 
a transition probability
from state $n_1, n_2$ to state $m_1, m_2$ as follows:
\begin{align}
&P_{n_1, n_2\to m_1, m_2}(t)\nonumber \\
&= \frac{n_1!n_2!}{m_1!m_2!} 
\sum_{n_\alpha = 0}^\infty \sum_{n_\beta = 0}^\infty 
\left( \frac{\alpha}{\alpha'} \right)^{n_\alpha}
\left( \frac{\beta}{\beta'} \right)^{n_\beta}
\phi(n_1,n_2,n_\alpha,n_\beta,t).
\label{eq_demo_final_dual}
\end{align}

\begin{figure}
\includegraphics[width=90mm,keepaspectratio]{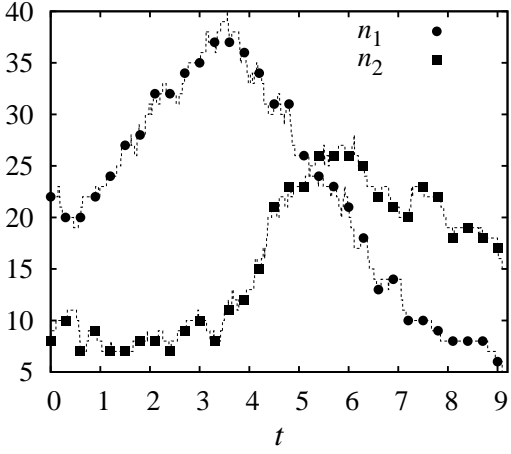}
\caption{
Measurement data.
Symbols denote partial observations with discrete time steps 
$\Delta \tau^\mathrm{obs} = 0.3$.
For comparison, the original processes
are plotted as dotted lines.
This artificial time-series data is generated
with the parameters 
$\alpha = 0.2,\beta = 0.02,\delta = 0.02$,
and $\gamma = 0.3$.
}
\label{fig_1}
\end{figure}

Notice that Eq.~\eqref{eq_demo_final_dual} has a form of 
the Taylor-type expansion around the origin.
Hence, the usage of the dual process corresponds to calculations of the expansion coefficients.
In addition, it is easy to calculate the derivatives with respect to the parameters
from the same $\phi(n_1,n_2,n_\alpha,n_\beta,t)$;
there is no need to perform additional time-evolution.

The derived formula for the transition probabilities in Eq.~\eqref{eq_demo_final_dual}
can be, for example, used in parameter estimation problems.
Assume that we have discrete-time observations
of time-series date, which are depicted in Fig.~\ref{fig_1}.
The observation-time interval is $\Delta \tau^\mathrm{obs} = 0.3$,
and the full time-series data are not available.
That is, 
only the observation values $n_1^{(i)}$ and $n_2^{(i)}$ 
at discrete time $\{t^{(i)} | \, i = 1, 2, \dots, N\}$ are available,
where $t^{(i+1)} - t^{(i)} = \Delta \tau^\mathrm{obs} = 0.3$,
and $N = 31$ in Fig.~\ref{fig_1}.

\begin{figure}
\includegraphics[width=86mm,keepaspectratio]{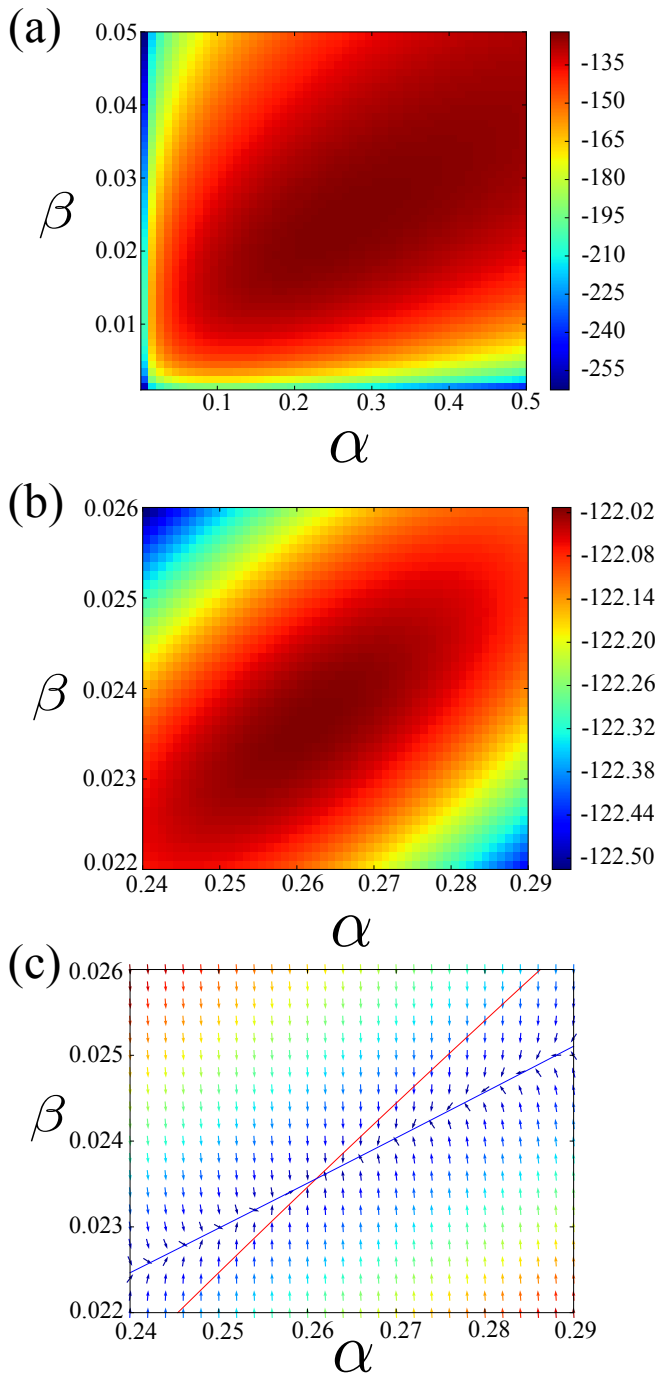}
\caption{
Heatmaps and nullclines for the log-likelihood function.
(a) and (b) show the heatmaps.
(b) is an enlarged one; note that the color profiles in (a) and (b) are different.
(c) shows the directions of derivatives and nullclines.
}
\label{fig_2}
\end{figure}

Assume that two parameters $\delta$ and $\gamma$ are known.
The other two parameters, $\alpha$ and $\beta$,
should be estimated from the time-series data.
In order to seek the parameters,
the following likelihood function is calculated;
\begin{align}
\mathcal{L}(\alpha,\beta)
= \prod_{i=1}^{N-1} 
\mathrm{Prob}(n_1^{(i+1)}, n_2^{(i+1)} | n_1^{(i)}, n_2^{(i)}; \alpha, \beta),
\end{align}
where $\mathrm{Prob}(n_1^{(i+1)}, n_2^{(i+1)} | n_1^{(i)}, n_2^{(i)}; \alpha, \beta)$
is the probability of the state change with parameters $\alpha$ and $\beta$.
Additionally, it is usual to calculate the following 
log-likelihood function,
\begin{align}
\ell(\alpha,\beta) = \log \mathcal{L}(\alpha,\beta),
\label{eq_ll}
\end{align}
instead of the original likelihood function.

In general, the log-likelihood function in Eq.~\eqref{eq_ll}
should be re-calculated for various values of $\alpha$ and $\beta$,
and the tasks need high computational costs.
In contrast, the formula in Eq.~\eqref{eq_demo_final_dual}
is suitable to depict the contour or heatmap for the log-likelihood function.
That is, values of the log-likelihood function
can be evaluated only from a time-integration of the extended dual process.
Hence, there is no need to repeat the time-integration for different parameters.
(Here, assume that the orders of scales of $\alpha$ and $\beta$
are previously known, and the settings $\alpha' = 1.0$ and $\beta' = 0.1$
are used in Eq.~\eqref{eq_demo_final}.)
Additionally, the information of the derivatives of the log-likelihood functions
is easily obtained from Eq.~\eqref{eq_demo_final_dual};
there is no need to perform additional time-integrations.

In order to demonstrate the usage of the duality relations,
the numerical calculations
for heatmaps and nullclines for the log-likelihood functions
are performed as follows.
Both Eqs.~\eqref{eq_demo_master_equation} and \eqref{eq_demo_final}
are coupled ordinary differential equations,
and the numbers of the equations are, in principle, infinite.
Usually, the time-integration needs a finite cut-off for states;
here, only states with $0 \sim 69$ are considered for each particle.
I checked that the finite cut-off is enough for the truncation
of the summation in Eq.~\eqref{eq_demo_final_dual}.
For the time-integration for the dual processes,
the usual 4th-order Runge-Kutta method is employed.
Figures~\ref{fig_2}(a) and (b) are the results for the heatmap;
Fig.~\ref{fig_2}(b) is the enlarged one of Fig.~\ref{fig_2}(a).
Figure~\ref{fig_2}(c) shows the directions of derivatives of the log-likelihood functions,
and nullclines.
Again, note that there is no need to perform additional time-integrations here;
only the same solutions for the extended dual process are repeatedly re-used
to depict all Figs.~\ref{fig_2}(a), (b), and (c).


\begin{figure}
\includegraphics[width=70mm,keepaspectratio]{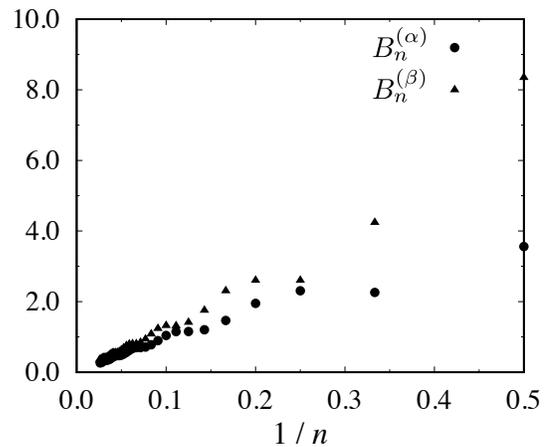}
\caption{
Plots of the quantities defined in 
Eqs.~\eqref{eq_bn_for_alpha} and \eqref{eq_bn_for_beta} (up to $n=30$.)
When numerical evaluation for larger $n$ cases is performed,
the intercept with $1/n = 0$ gives the radius of convergence.
The plots correspond to the cases
with ${n_1 = 37}, {n_2 = 11}, {m_1 = 36}, {m_2 = 12}$.
For $B_n^{(\alpha)}$, $\beta' = 0.1$ and $\beta = 0.0236$ are used;
$\alpha' = 1.0$ and $\alpha = 0.261$ are used for $B_n^{(\beta)}$.
}
\label{fig_3}
\end{figure}

Finally, I give some comments for the radius of convergence of the power series 
in Eq.~\eqref{eq_demo_final_dual}.
Equation~\eqref{eq_demo_final_dual} has two variables,
and we here introduce the following coefficients:
\begin{align}
&c_{n}^{(\alpha)} 
\equiv
 \frac{n_1!n_2!}{m_1!m_2!} 
\sum_{n_\beta=0}^\infty
\left( \frac{\beta}{\beta'} \right)^{n_\beta}
\phi(n_1,n_2,n, n_\beta,t), \label{eq_conv_alpha_coeff}\\
&c_{n}^{(\beta)}
\equiv 
 \frac{n_1!n_2!}{m_1!m_2!} 
\sum_{n_\alpha=0}^\infty
\left( \frac{\alpha}{\alpha'} \right)^{n_\alpha}
\phi(n_1,n_2,n_\alpha, n, t).
\label{eq_conv_beta_coeff}
\end{align}
Note that $c_n^{(\alpha)}$ and $c_n^{(\beta)}$
depend on $n_1$, $n_2$, $m_1$, $m_2$, $\beta$, $\beta'$ and $t$;
the dependencies are not explicitly shown in $c_n^{(\alpha)}$ and $c_n^{(\beta)}$
for notational simplicity.
Then, the transition probability is written in the following
power-series with one variable:
\begin{align}
P_{n_1, n_2\to m_1, m_2}(t)
&=
\sum_{n = 0}^\infty 
c_{n}^{(\alpha)}
\left( \frac{\alpha}{\alpha'} \right)^{n} \label{eq_conv_alpha}\\
&=
\sum_{n = 0}^\infty 
c_{n}^{(\beta)}
\left( \frac{\beta}{\beta'} \right)^{n}.
\label{eq_conv_beta}
\end{align}
In order to discuss the radius of convergence of the power-series,
one may use the so-called Domb--Sykes plot \cite{Domb1957}.
Here, we use the method introduced by Mercer and Roberts \cite{Mercer1990}
because complicated patterns appear in $\{c_n^{(\alpha)}\}$ and $\{c_n^{(\beta)}\}$.
In the Mercer--Roberts method,
the following quantities are calculated:
\begin{align}
(B^{(\alpha)}_n)^2 &= 
\frac{c_{n+1}^{(\alpha)} c_{n-1}^{(\alpha)} - (c_n^{(\alpha)})^2}
{c_n^{(\alpha)} c_{n-2}^{(\alpha)} - (c_{n-1}^{(\alpha)})^2}, \label{eq_bn_for_alpha}\\
(B^{(\beta)}_n)^2 &= 
\frac{c_{n+1}^{(\beta)} c_{n-1}^{(\beta)} - (c_n^{(\beta)})^2}
{c_n^{(\beta)} c_{n-2}^{(\beta)} - (c_{n-1}^{(\beta)})^2}, \label{eq_bn_for_beta}
\end{align}
for $n = 2,3,4,\dots$.
As discussed in \cite{Mercer1990},
for example, the reciprocal of the radius of convergence, $1/r$, for Eq.~\eqref{eq_conv_alpha}
is given as the intercept with $1/n = 0$
when we plot $B_n^{(\alpha)}$ versus $1/n$.
Figure~\ref{fig_3} shows the analysis;
the plots correspond to the case
with $n_1 = 37, n_2 = 11, m_1 = 36, m_2 = 12$,
$\alpha' = 1.0$, $\alpha = 0.261$,
$\beta' = 0.1$, and $\beta = 0.0236$.
(Other parameters give the similar behaviors.)
The results in Fig.~\ref{fig_3} would not be enough
to obtain the precise values of the radius of convergence;
the coefficients with larger $n$ in the power-series are needed,
but there are limitations of the memory capacity in the Runge-Kutta methods and numerical precision.
However, as shown in Fig.~\ref{fig_3},
the intercepts with $1/n=0$ seem to take near-zero values,
and the radius of convergence would be large enough.

\section{Concluding remarks}
\label{sec_conclusion}

As shown in the present paper,
the duality relations enable us to reduce
the repeated time-integrations for various rate constants.
This feature is applicable, for example,
to obtain heatmaps for the log-likelihood functions;
compared to the direct time-integrations for the original system,
only one time-integration for the extended dual process is enough,
as demonstrated in the present paper.
In addition, the derivatives of the log-likelihood functions
can also be obtained easily by using the same numerical solutions
for the extended dual process.
Of course, it is also possible to calculate the Jacobian matrix, the Hessian matrix, and so on.

It should be noted here
that the number of random variables in the extended dual processes
is larger than the original one.
Hence, the computational cost for one time-integration 
becomes larger than the original one.
However, if we can perform the time-integration for the extended dual process
with reasonable computational costs,
the numerical results can be repeatedly re-used,
which will finally reduce the whole computational costs.

Here, the current limitations of the usage of the duality relations
should be stated.
\begin{enumerate}
\item As discussed above, 
the usage of the duality relations corresponds to the Taylor-type expansion. 
Hence, we must pay attention to the convergence.
In practice, it is preferable that the expanded variables take values smaller than $1$.
In order to avoid this convergence issue, 
it is useful to use the replacements (variable transformations) in Sec.~III.
\item In the demonstration, the coupled ordinary differential equations
for the dual process should be solved numerically.
When the numbers of parameters in the expansion are large,
it becomes impossible to solve the coupled ordinary differential equations
(the curse of dimensionality.)
Hence, at this stage,
the current approach is suitable to see the behavior of the log-likelihood functions
with a few varying variables.
\item In order to treat many variable cases, 
one may wonder if the Monte Carlo approach is available.
It is, in principle, true; 
although the time-evolution equation for the dual process does not correspond to
a stochastic process in general,
the stochastic nature can be recovered by using more extensions (see \cite{Ohkubo2013b}.)
Actually, as for the duality between the stochastic differential equations
and the birth-death processes, the Monte Carlo approach has already been employed \cite{Ohkubo2015}.
However, the Taylor-type expansion needs sometimes solutions with high precisions,
and the Monte Carlo approach is still time-consuming.
The usage of the importance sampling may avoid this problem;
this is beyond of the scope of the present paper, and under investigation.
\end{enumerate}

At the present moment, the duality relations are useful
to investigate cases with a few varying variables.
In future works, it is important to develop
approximation methods or efficient numerical methods 
for the extended dual processes.

\vskip 2mm

\section*{Acknowledgement}
This work was supported in part by MEXT KAKENHI 
(Grants no. 25870339 and 16K00323).

\end{document}